\documentclass[pre,aps,showpacs]{revtex4} % showkeys

\usepackage{amsmath}
\usepackage{graphicx}% Include figure files
\usepackage{dcolumn}% Align table columns on decimal point
\usepackage{bm}% bold math

\begin{document}

\title{Physics of evolution: selection without fitness}
%{CLT\&MEP}

\author{Stefan Thurner$^{1,2,*}$ and Rudolf Hanel$^{1}$}
%\email{thurner@univie.ac.at} 

\affiliation{
$^1$ Complex Systems Research Group; HNO; Medical University of Vienna; 
W\"ahringer G\"urtel 18-20; A-1090; Austria \\
$^2$ Santa Fe Institute; 1399 Hyde Park Road; Santa Fe; NM 87501; USA\\
$^*$ thurner@univie.ac.at
}

%\date{Version \today}

\begin{abstract}
Traditionally evolution is seen as a process where from a pool of possible variations of a population 
(e.g. biological species or industrial goods) a few variations get selected which survive and 
proliferate, whereas the others vanish. 
Survival probability is typically associated with the 'fitness' of a particular variation. 
In this paper we argue that the notion of fitness is an {\em a posteriori} concept, in the sense 
that one can assign higher fitness to species that survive but one can generally not derive or 
even measure fitness -- or fitness landscapes -- {\it per se}.  
For this reason we think that in a 'physical' theory of evolution such notions should be avoided. 
In this spirit, here we propose a random matrix model of evolution where selection mechanisms 
are encoded in the interaction matrices of species. We are able to recover some key facts 
of evolution dynamics, such as punctuated equilibrium, i.e. the existence of intrinsic  large 
extinctions events, and, at the same time, 
periods of dramatic diversification, as known e.g. from fossil records.  
Further, we comment on two fundamental technical problems of a 'physics of evolution', 
the non-closedness of its phase space and  the problem of co-evolving boundary  conditions, 
apparent in all systems subject to evolution.
\end{abstract}

\pacs{
87.23.Ke, 
87.10.e,
87.23.Cc,
02.10.Ox
}% PACS

\maketitle

\section{Introduction}

The  quantitative science of evolution can look back on a series of models dealing with the evolution of 
biological systems and the dynamics of technological innovation. In this context maybe the most 
important ones are the models of  Kauffman (see e.g. \cite{skauff})  on biological and Arthur \cite{barthur} 
on technological evolution. A traditional way to treat dynamical systems of evolution 
is to use high dimensional catalytic network equations \cite{farmer,stadler}, the 
Lotka-Volterra equations and  the Hypercycle \cite{eigen} being famous examples. 
Recently 
%in 
linear versions of these catalytic network equations have been shown to demonstrate 
the influence of topological aspects of the underlying network topology on the resulting 
population dynamics \cite{jainkrishna}. There the existence of highly populated phases have been 
demonstrated to coincide with the appearance of closed directed feedback loops  
in the network, the so-called autocatalytic cycles.
On the subject of  extinction  dynamics, which constitutes an important branch in understanding evolution, 
several specific models have been brought forward lately \cite{baksneppen,solemanrubia,newman,raup,klimek}.

Of fundamental importance in all of the above models is the notion of 
{\em fitness}. Since the times of Darwin fitness is used  as a method for selection, which is 
fundamental to evolution. Fitness of an entity is a function  depending on a large number of parameters, 
describing both the state of the entity  and the state of its environment. Due to the high dimensional 
parameter space one often talks about fitness landscapes. These landscapes may be rather complicated 
and hard to optimize. 
As useful as it may be for an intuitive feeling of how selection works, 
the notion of fitness is clearly an {\em  a posteriori} concept. An entity is fit if it 
had survived  and proliferated well, and unfit if it went extinct. The same is true for 
the 
%important 
concept of  {\em ecological niches}. 
Space that never got occupied by entities will not count as a niche.
%A niche is whatever got occupied by 
%entities. 
%From a physics point of view 
These concepts could be seen as ambiguous, and anthropomorphic. 
%Further, 
Fitness and niches can not be measured in reality in much the same way as utility functions can usually not be measured in 
economics. 

\begin{figure}[t]
\includegraphics[width=10.5cm]{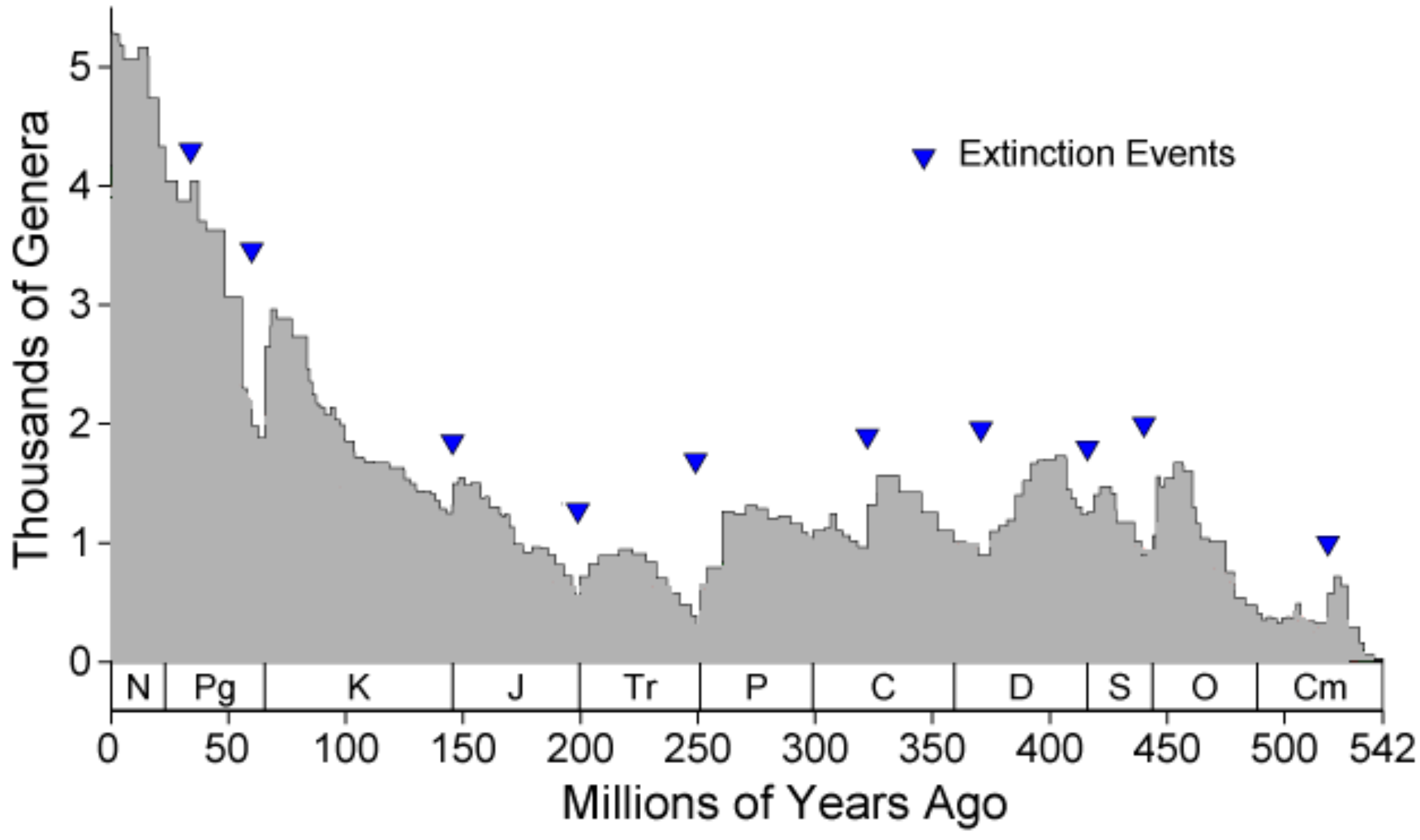}
\caption{
Number of biological genera over time. Clearly visible are the mass extinctions marked by triangles, 
often followed by periods of massive diversification (punctuated equilibrium).  Time runs from right to left.
Data after \cite{wiki}.
}
\label{fig1}      
\end{figure}

If there exists something like a physics of evolution it would be necessary to phrase models without 
using anthropomorphic concepts and, at the same time, provide a framework  abstract enough to
capture the universality behind e.g. technological and biological evolution in a unified way. 
This universality manifests itself in the fields of biological evolution, industrial innovation, socio-dynamics, economy, finance,
opinion formation, ecological dynamics, e.g. food-webs, and history.   
In the following we would like to avoid these notions in reasoning about evolution all together.
We propose a model able to explain several facts about evolutionary dynamics -- such as the 
existence  of intrinsic large extinctions events and 
booms 
%upsurges 
of diversity of species over time (punctuated equilibrium), see Fig. \ref{fig1} --  
based on three 
guiding principles:
{\em low resolution},  {\em parsimony} and {\em maximum ignorance}. 
Resolution of detail is limited in many evolutionary time series. Practically 
one can often only resolve and understand the dynamics of coarse variables as, for example, 
the diversity of the system, i.e. how many different species are actually evolving together. 
Parsimony (Occam's Razor) calls for simplistic models based on a minimum of resolvable 
explanatory parameters. Finally, maximum ignorance suggests to model  
interactions which are not known in detail on the basis of random matrices (tensors) as  
introduced by Wigner \cite{wigner}.

Generally speaking, evolution can be understood as an iterative three-step process.
{\em Step 1:} a {\em new thing} comes into existence within a given {\em environment}, 
usually on the basis of existing things by means of substitution and combination processes.
Environment is defined as the set of all existing things in the  system. 
{\em Step 2:} The new thing has the chance to interact with the environment. As a result of this 
interaction the thing  gets selected (survives) or de-selected (destroyed, suppressed).
{\em Step 3:} In case the new thing  gets selected  within this environment it becomes part of
the environment for other things, thereby changing the conditions for both - existing things and  
future new things yet to arrive. The new thing, if selected, has the chance to 'proliferate'  in terms of 
substitution and combination processes with itself or other existing things.
  
It is clear that such a process is  far from being 
of Newtonian nature as usually dealt with in physics.
The most obvious problem arises from the innovations happening in the course of the dynamics:  
the phase space of the system is not closed. 
Whenever  a satisfactory set of variables is found that describes the history of the system well, 
at some point this set of variables becomes insufficient to describe states of the system in the future. 
For example, before television was invented history of mankind can be understood without 
television, whereas it gets increasingly impossible to understand the history of man after the 
invention without taking the secondary effects of television on society into account. 
The second major problem  is that of co-evolving boundary conditions, 
which all evolution systems share. As the system evolves 
(new things come into being) the boundary conditions (interaction possibilities with others) 
constantly change. This makes such systems extremely hard to phrase in terms of differential equations. 
%This fact also might also make it hard to think of an existence of a reasonable physics of evolution. 
Since it is impossible to forecast the invention of entities and new variables associated with them, 
it seems reasonable, as a starting point, to model their respective interactions based on random interactions.

As stated above new things usually come into existence through a process of (re-) combination or substitution 
of already existing things. Examples range from sexual reproduction to the assembly of an ipod from its components.  
Thus models for evolutionary systems minimally require  two components: 
{\em entities} (things, species, products, individuals, etc.)  and 
{\em rules} how these entities can be used in combination with each other to produce new things (combination and substitution).  For example there is a rule that the combination of 
hydrogen and  oxygen can form water, or there is a rule that  a male and female chicken  can become parents of  baby chicken. 
However there is no rule that a fish and a dog can be the parents of a chicken, nor is it possible  
to combine two blocks of U 235  into one big block of uranium. The set of all combination/substitution rules 
we keep in a 'rule table' denoted as $\alpha$ in the following.

Suppose we characterize the relative frequency of entity $i$ by $x_i\geq 0$ and denote
the rule whether entity $i$ can be produced from entities $j$ and $k$
by $\alpha_{ijk}$. If no such production rule exists, then $\alpha_{ijk}=0$ and else $\alpha_{ijk}>0$, when such a rule exists. 
$\alpha_{ijk}$ represents the rate with which $j$ and $k$ can produce $i$. For simplicity
one can view $\alpha$ as Boolean, i.e. $\alpha\in\{0,\,1\}$. If $\alpha_{ijk}>0$ we call $(j,k)$ a {\em creative pair} producing (or 'pointing' to) $i$. 
%This production of new things on the basis of creative pairs is indicated in Fig. (\ref{fig2}).
The dynamics of the relative frequencies of the entities can be expressed by a network equation of the type
\begin{equation}
\dot{x}_{i}= \sum_{j,k}  \alpha_{ijk}x_j x_k -x_{i}\Phi \quad , \quad 
       \Phi= \sum_{i,j,k}\alpha_{ijk}x_j x_k \quad.
\label{networkeq}
\end{equation}
The quadratic term suggests that $i$ forms proportional to the abundance of $j$ and $k$ weighted
by the rate $\alpha_{ijk}$. The $\Phi$ term %only 
assures that the abundance of entities is proportional to the
%measured in 
relative frequencies $x_i$, with
$\sum x_i = 1$.

\begin{figure*}[t]
\includegraphics[width=3cm]{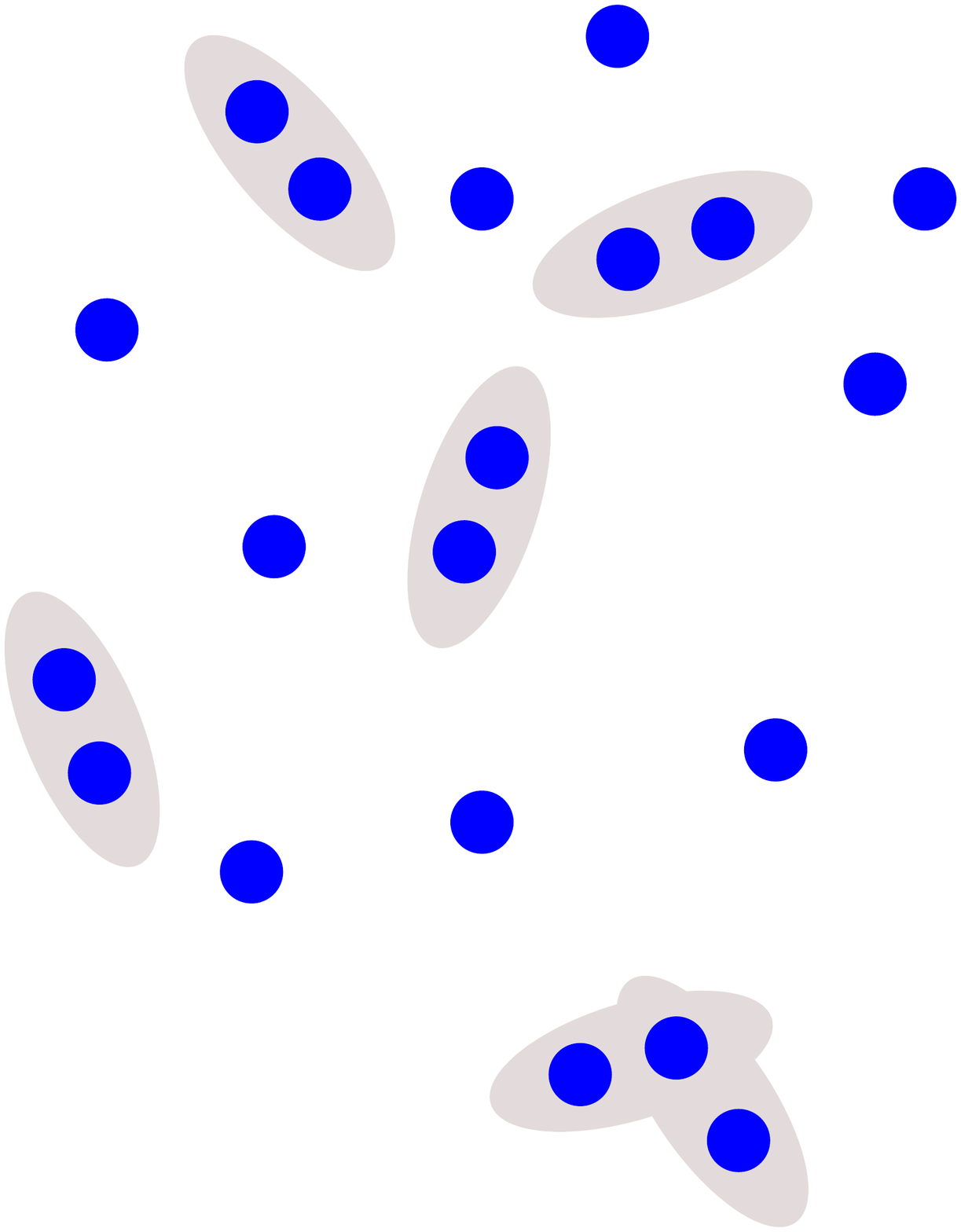}
\includegraphics[width=3cm]{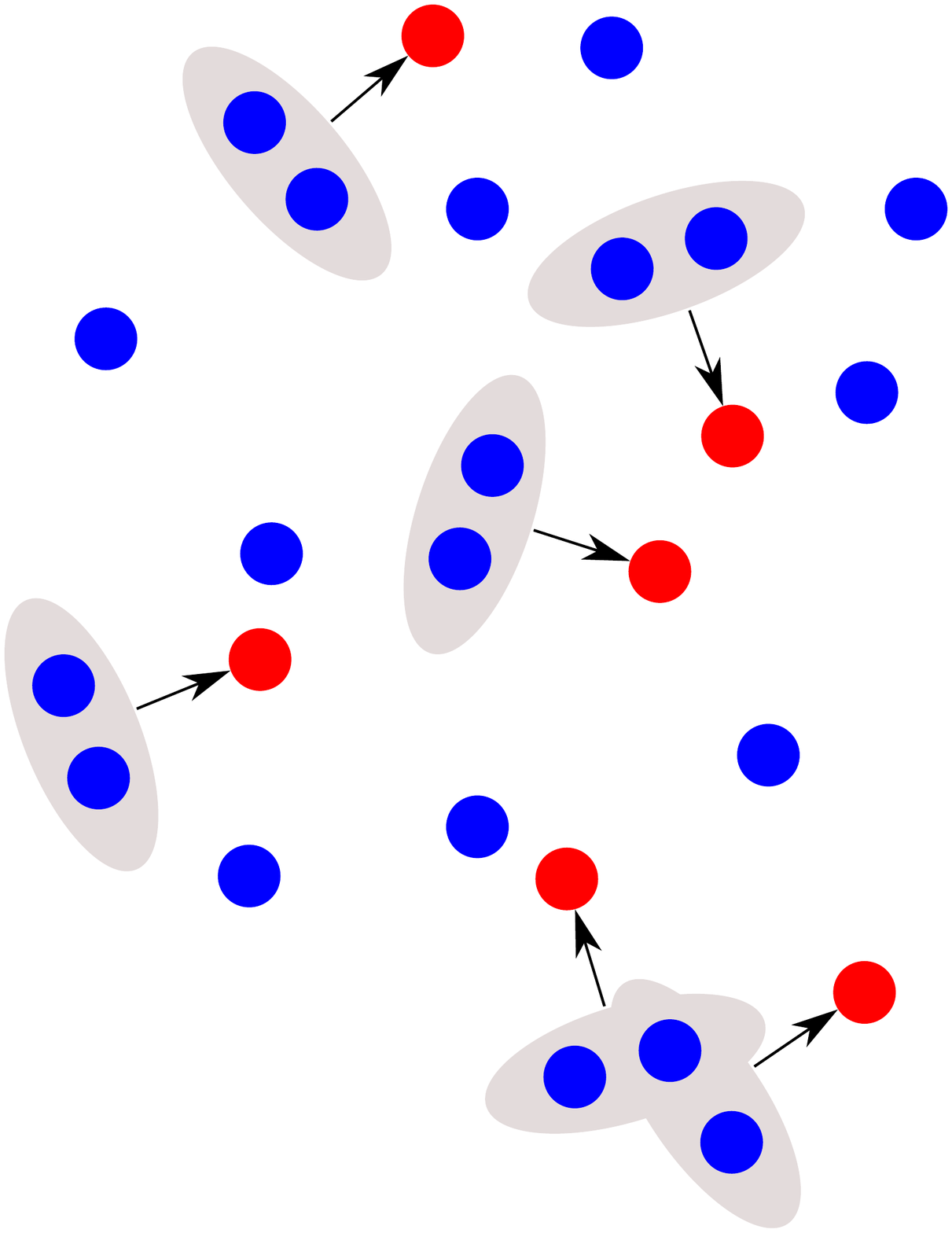}
\includegraphics[width=3cm]{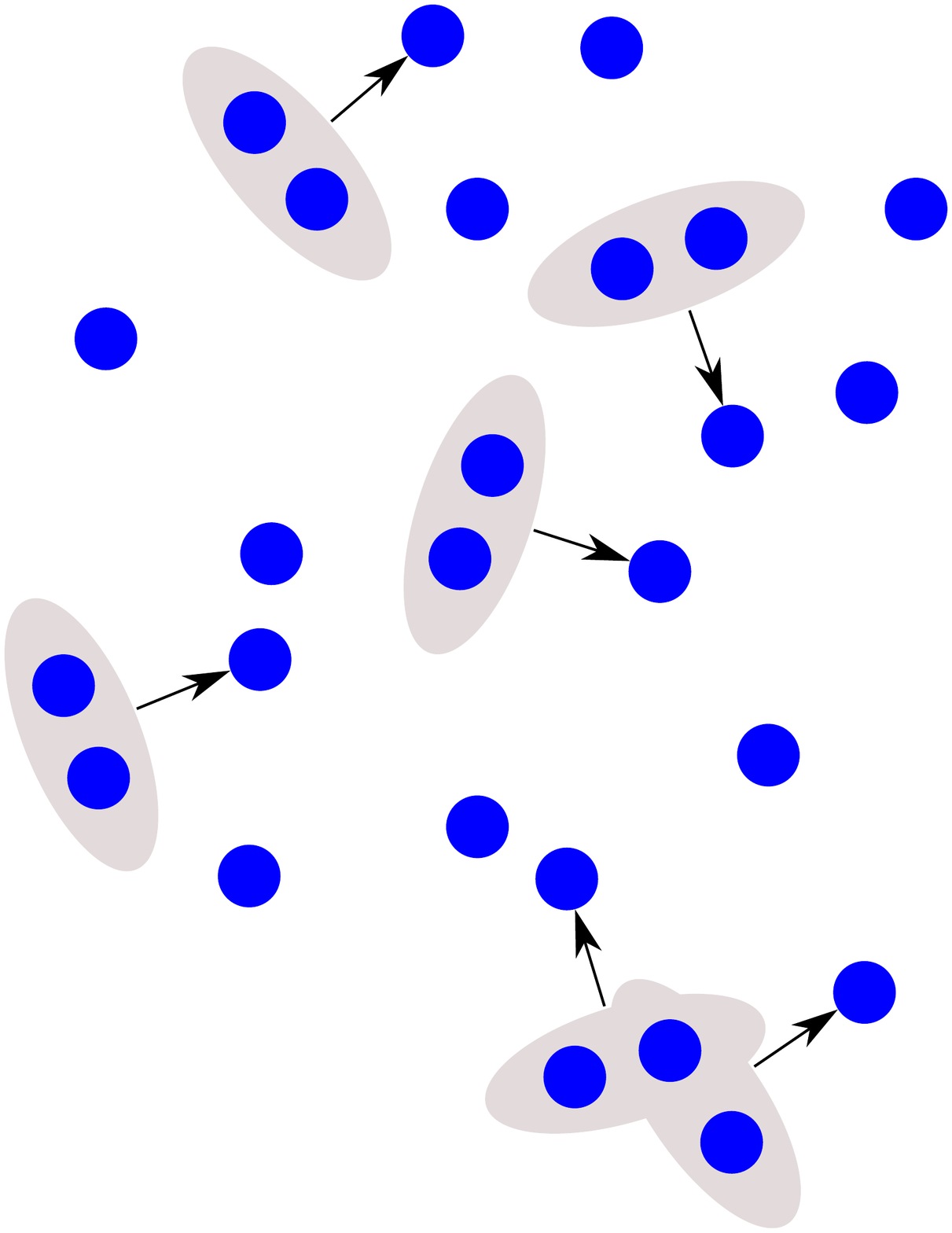}
\includegraphics[width=3cm]{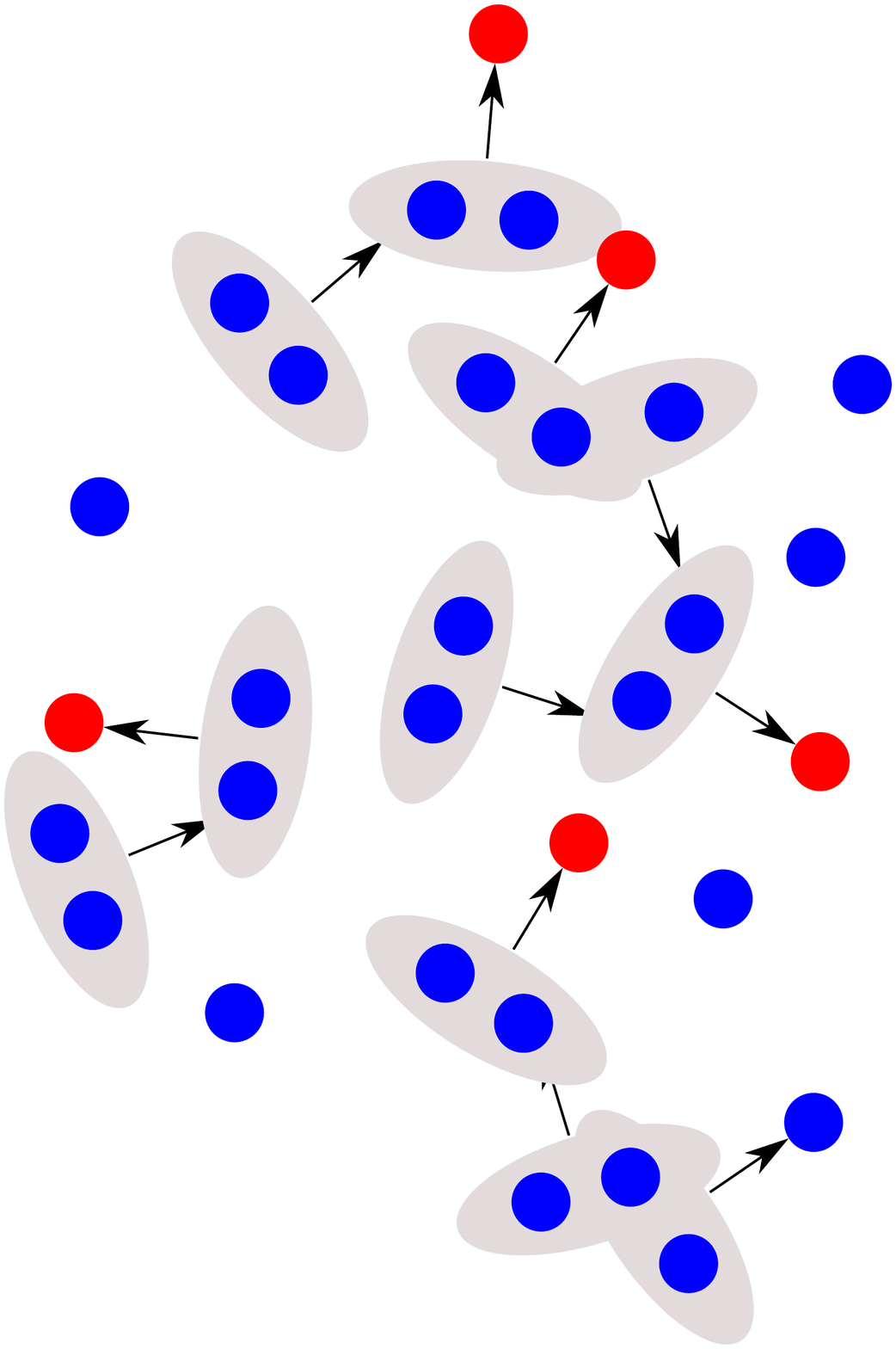}
\caption{
From left to right: within some initial set of entities (blue dots) there exist several creative pairs, 
indicated by gray ovals. In the next timestep these pairs produce new entities (red dots). 
The new entities become part of the environment (red dots turn blue).
With the new entities new productive pairs can be formed to again produce new entities.
}
\label{fig2}      
\end{figure*}

\begin{figure*}[t]
\includegraphics[width=10cm]{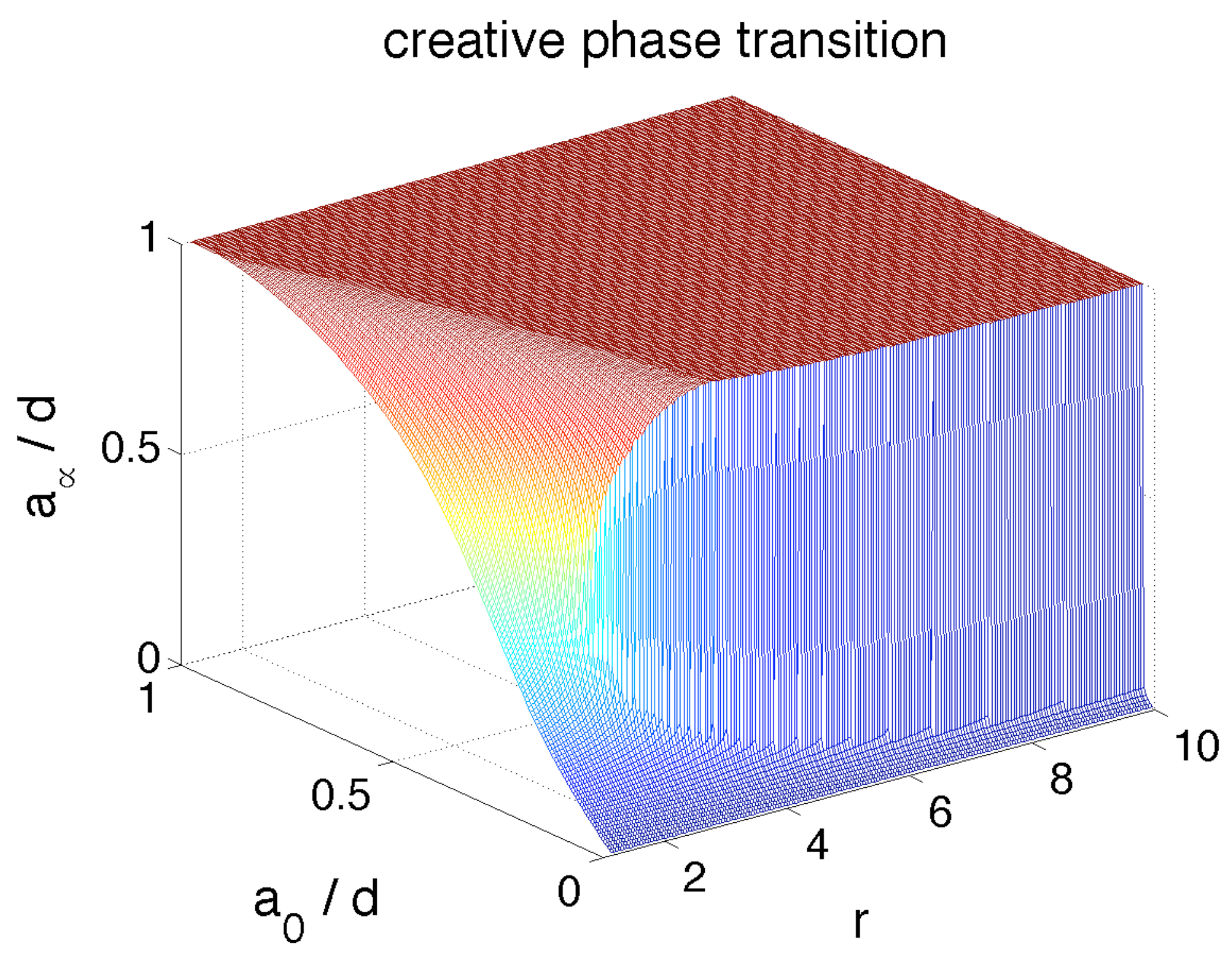}
\caption{
Final diversity $a_\infty$ plotted over the initial diversity $a_0$ and the creative pair density $r^+$
displaying a van der Waals like phase transition from low to  high final diversity.
}
\label{fig3}      
\end{figure*}

Let us assume that the dimension of the system, denoted as $d$, i.e. the  number of all possible entities $i$, is extremely large, but finite. 
If we know all $i$ and $\alpha_{ijk}$ we could specify an initial condition $x(t=0)$ and solve the
differential equation (\ref{networkeq}). However, neither $d$ nor $\alpha$ is known in general. So how can one proceed? In previous 
work \cite{htk1,htk2} we suggested the following combinatorial approach.
Suppose we knew the number of productive pairs $N^{+}=\sum_{i,j,k}\theta(\alpha_{ijk})$ (where $\theta(x)=1$ 
for $x>0$ and $\theta(x)=0$ otherwise) in the system and define 
the creative pair density $r^+=N^+/d$, which is the average number of productive pairs leading to one specific product. 
Suppose further that the topology of the productive catalytic network is random (maximum ignorance), 
then one can to use combinatorial arguments to map the catalytic network equation (\ref{networkeq}) 
onto the recurrence relation, \cite{htk1}  
\begin{equation}
\begin{array}{l}
 a_{t+1}=a_{t}+\Delta a_{t} \\
 \Delta a_{t+1}=r^+\left(1-a_{t+1}\right)
 \left(a_{t+1}^{2}-a_{t}^{2}\right) \quad,
\end{array}
\label{growth}
\end{equation} 
where $a_t\equiv \sum_{i}\theta(x_i(t))/d$ is the relative diversity of the system at time $t$, 
i.e. the fraction of possible entities with non-zero abundance.  Here the  
initial condition is $a_{0}$ and $a_{-1}=0$ by definition; the rule density is $r^+$. 
For a schematic iterative view of a dynamical increase of diversity see Fig. \ref{fig2}.
It is remarkable that this equation does not explicitly depend on $d$ any more and therefore is valid for arbitrarily  
large $d$. The increment $\Delta a$ in Eq. (\ref{growth}) can %easily 
be understood. 
%First of all 
$r^+ a_t^2$ is proportional to
the number of productive pairs in $a_t$. However, we have to subtract the productive pairs that have already been considered in the
last time-step so that at time $t$ only $r^+(a_t^2-a_{t-1}^2)$ new pairs have to be taken into account. Of these pairs only a fraction
$(1-a_t)$ can be expected to produce something that does not already exist. The resulting update equation Eq. (\ref{growth}) 
can be solved asymptotically by analytical means \cite{htk1} and shows that the system has a phase 
transition in the final diversity $a_\infty$. 
If $r^+>r^+_{crit}$, with a critical value $r^+_{crit}\sim 2$, then there exists a critical value
$a_{crit}(r^+)$ where there occurs  a discontinuous jump in the final diversity from low to high final diversity. 
The transition is equivalent  to the van der Waal's gas phase transition. This phase transition is shown 
in Fig. (\ref{fig3}) where the final diversity $a_\infty$ is plotted over the initial diversity $a_0$ and the creative pair density $r^+$.
We have analyzed the stability of the fully populated state under external suppression
of entities (initial defects) which also leads to scale-invariant update equations for the propagation of secondary defects.
Here also a transition from almost linear behavior with respect to the initial defects to a break-down
of the complete system can be demonstrated, \cite{htk2}.

\begin{figure*}[t]
\includegraphics[width=3cm]{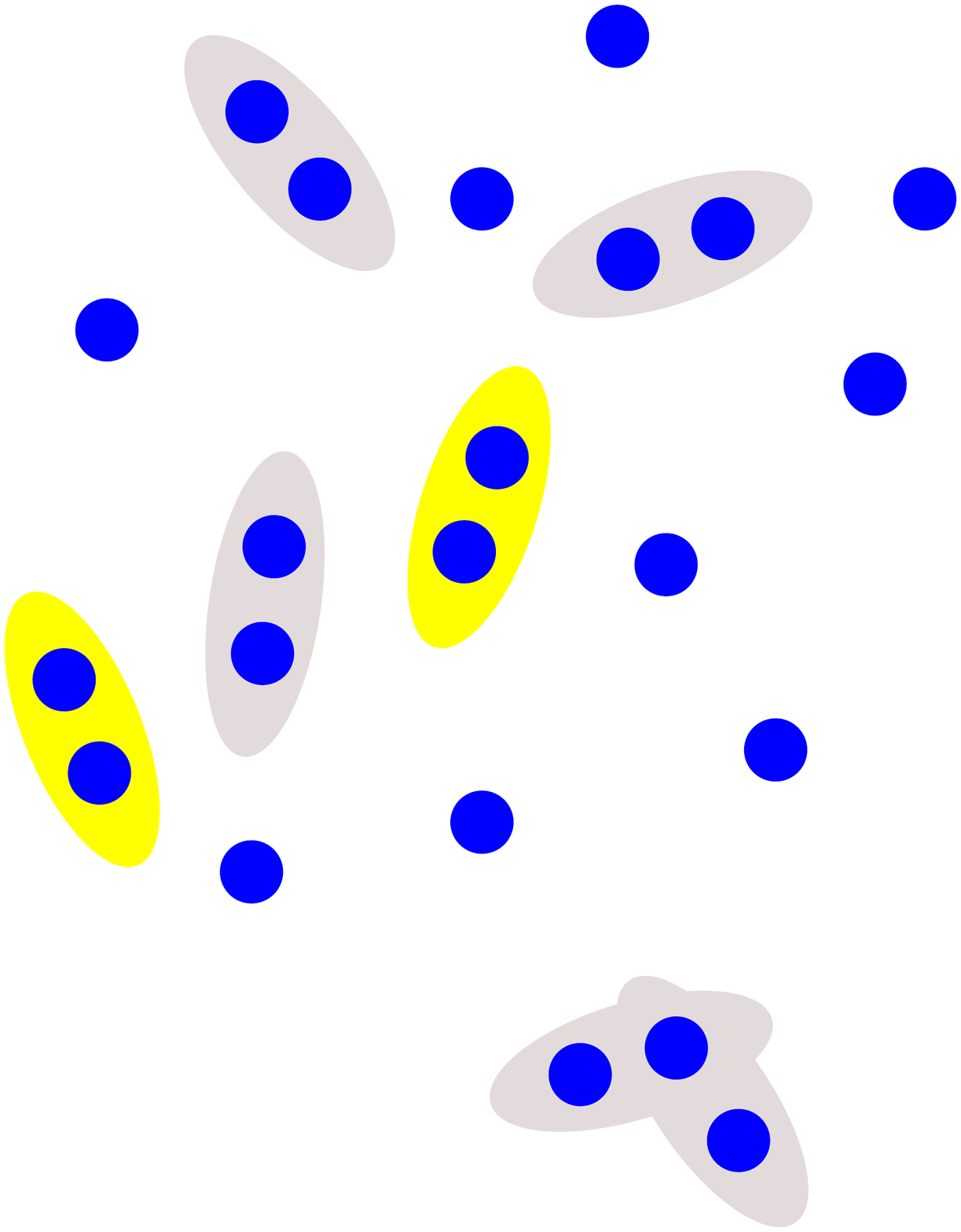}
\includegraphics[width=3cm]{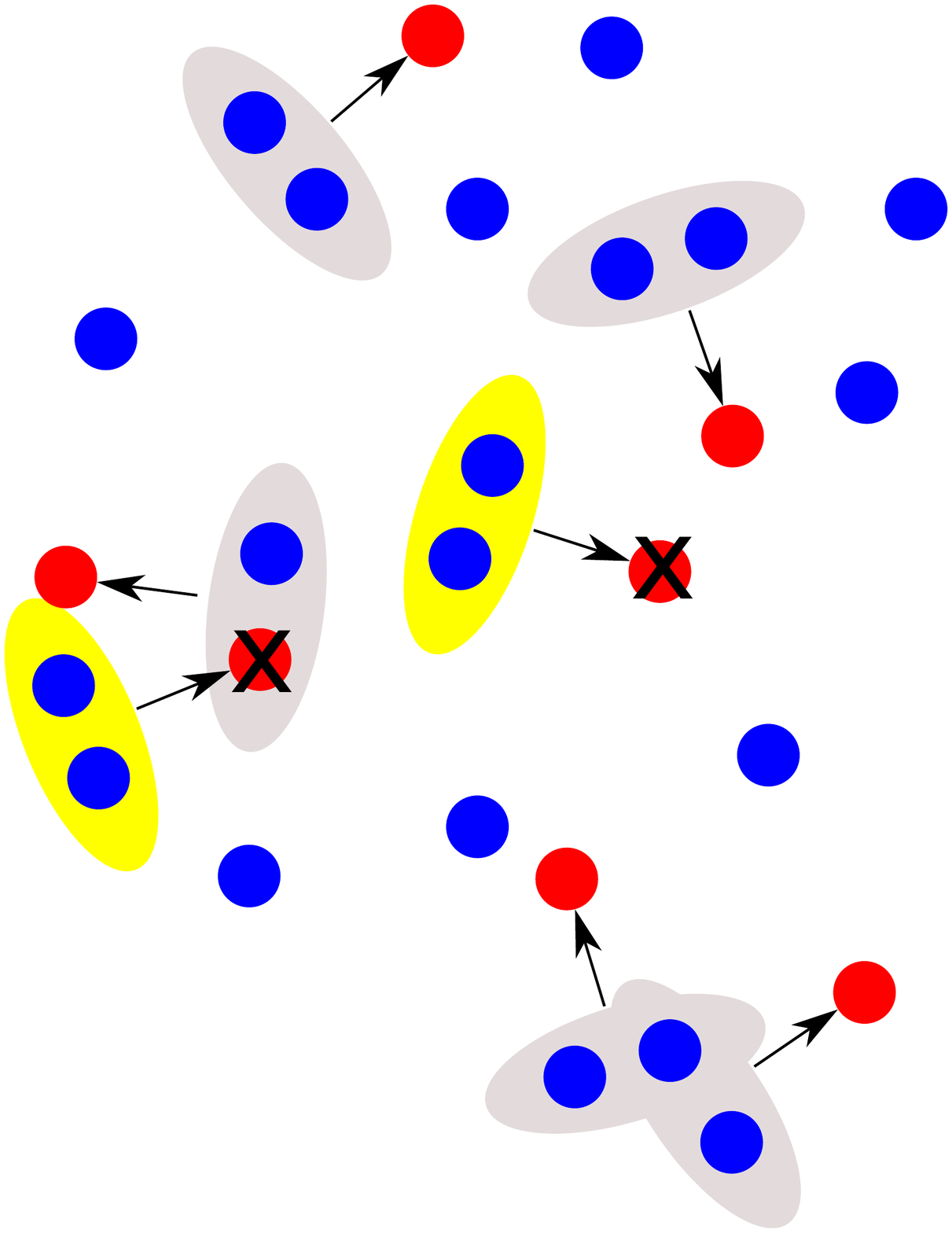}
\includegraphics[width=3cm]{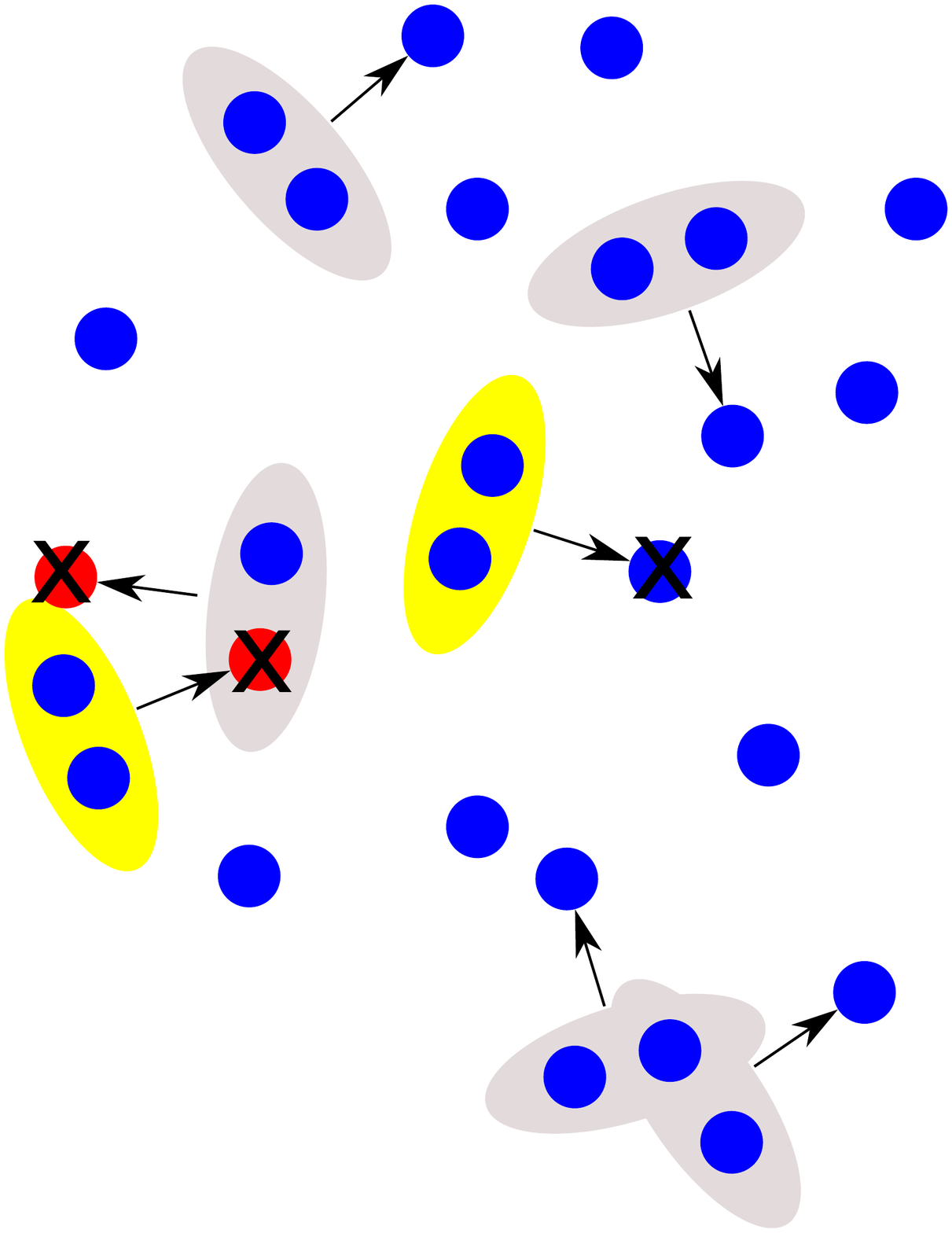}
\caption{
Left: Within some initial population of entities (blue dots) we find  creative pairs (gray ovals) 
as well as destructive pairs (yellow ovals). In the next timestep creative pairs produce new elements 
(red dots), suppressive pairs annihilate existing elements (crossed red dots). The resulting population 
with its potentially new pairs serve as a new initial population for the next timestep. 
}
\label{fig4}      
\end{figure*}

\section{An evolution model without fitness}

The above creative catalytic network dynamics does not yet constitute a model for evolution. 
The selection mechanism  as an intrinsic part of an evolution  system is missing.
The task is how to introduce selection without changing the set-up and without
reference to anthropomorphic {\em a posteriori} concepts like fitness, niches, selection pressure, etc. 

A straight forward way to achieve this, is to include suppressive pairs $(j,k)$ with $\alpha_{ijk}<0$ such that the
$(j,k)$ suppresses the existence of $i$. Rule table $\alpha_{ijk}$ now encodes creation and suppression,
$r^+$ remains the density of productive rules, and the density of suppressor rules, denoted by 
$r^-$, is defined analogously as  $r^-=N^{-}/d$, with  
$N^{-}=\sum_{i,j,k}\theta(-\alpha_{ijk})$.
A schematic view of the model is presented in Fig. \ref{fig4}. 

The influence of suppressors on the final diversity of the system can be estimated analytically. 
For instance, if $a_\infty$ is the final diversity for the case without suppressors, then the asymptotic  
diversity of a system with a suppressor density of $r^-$  can be derived on the basis of 
combinatorial arguments to be $\gamma a_\infty$, where 
\begin{equation}
\gamma=\frac{\sqrt{1+4r^-a_\infty^2}-1}{2r^-a_\infty^2}\leq1 \quad. 
\label{suppression-magnitude}
\end{equation}
This result has been shown to coincide reasonable well with simulations. 
Unfortunately it turns out that the diversity 
dynamics with suppressors is rather 
%resilient 
resistant to a more detailed 
analytical analysis and a numerical version of the model has to be implemented on the 
basis of the parameters $r^+$, $r^-$ and $a_0$ to learn more about the details of the dynamics 
of such systems.

\subsection{Implementation of the model}

We have implemented the most simple version of the above model.  
We consider only binary states $x_i\in\{0,1\}$, an element is present or absent,  and a rule table with entries 
$\alpha_{ijk}\in\{-1,0,1\}$. 
%The dimension of the system is $d=500$.
First we sample $r^+ d$ random triples $(i,j,k)$ and assign them a value $\alpha_{ijk}=1$. 
On the set of remaining triples we do the same for $r_-$ and assign $\alpha_{ijk}=-1$.  
All remaining entries in $\alpha$ are zero. 
%If the same is chosen ???? 
Next we randomly sample an initial condition such that exactly 
$a_0 d$ components in the initial species  vector $x(t=0)$ are one;  all others are zero. 
The elements $x_i(t)$ are updated to time $t+1$ in a random sequential order. 
For each node $i$ we count the number $n^+$ and $n^-$ of productive and suppressive pairs pointing to 
node $i$, respectively. If $n^++n^->0$ the node will be set to 1 with a probability $p_1=n^+/(n^++\mu n^-)$, and to 0 with probability $p_0=1-p_1$.
Here $\mu$ is a parameter that specifies the relative strength of suppressive pairs  over creative pairs. For example, 
if $\mu=1$ and $n^+=n^-$ then the chance for an entity $i$ to be created or deleted is $0.5$. 
If neither a creative pair nor a destructive pair is pointing at node $i$, i.e. $n^++n^-=0$ 
then $i$ does not get actively suppressed but also not actively produced. 
In this case $i$ will continue to exist in the next time step with a probability
$\lambda$ (sustain rate), in other words, entities that are not actively produced decay with a rate $1-\lambda$.
%This decay is happening for all entities except for the initially present  elements. 
At this point we can decide to protect the initial condition $x(0)$ or the initial diversity $a_0$, 
i.e. once the diversity $a(t)$ has dropped down to the level of $a_0$ no further entity can be eliminated 
in the update. However, we found that both types of enforcing a minimum diversity lead to 
very similar characteristics of the dynamics. 
This protection of entities models e.g. sources of renewable goods which are neither subject to selection nor to decay.

The dynamics of species in vector $x(t)$ can be seen as a point in phase space.
Two consecutive points in time have shifted by an 'angle' $\delta(t)$
\begin{equation}
\cos \delta (t) = \frac{ \vec x(t)  \vec x(t-1) }{|\vec x(t) | |\vec x(t-1)
| }\quad,
\label{radspeed}
\end{equation}
which can be regarded as a potential measure for the relative change of the population of entities 
and thus for the size of creative/destructive effects within this timestep.

\begin{figure*}[htp]
\begin{tabular}{ccc}
\includegraphics[width=5.4cm]{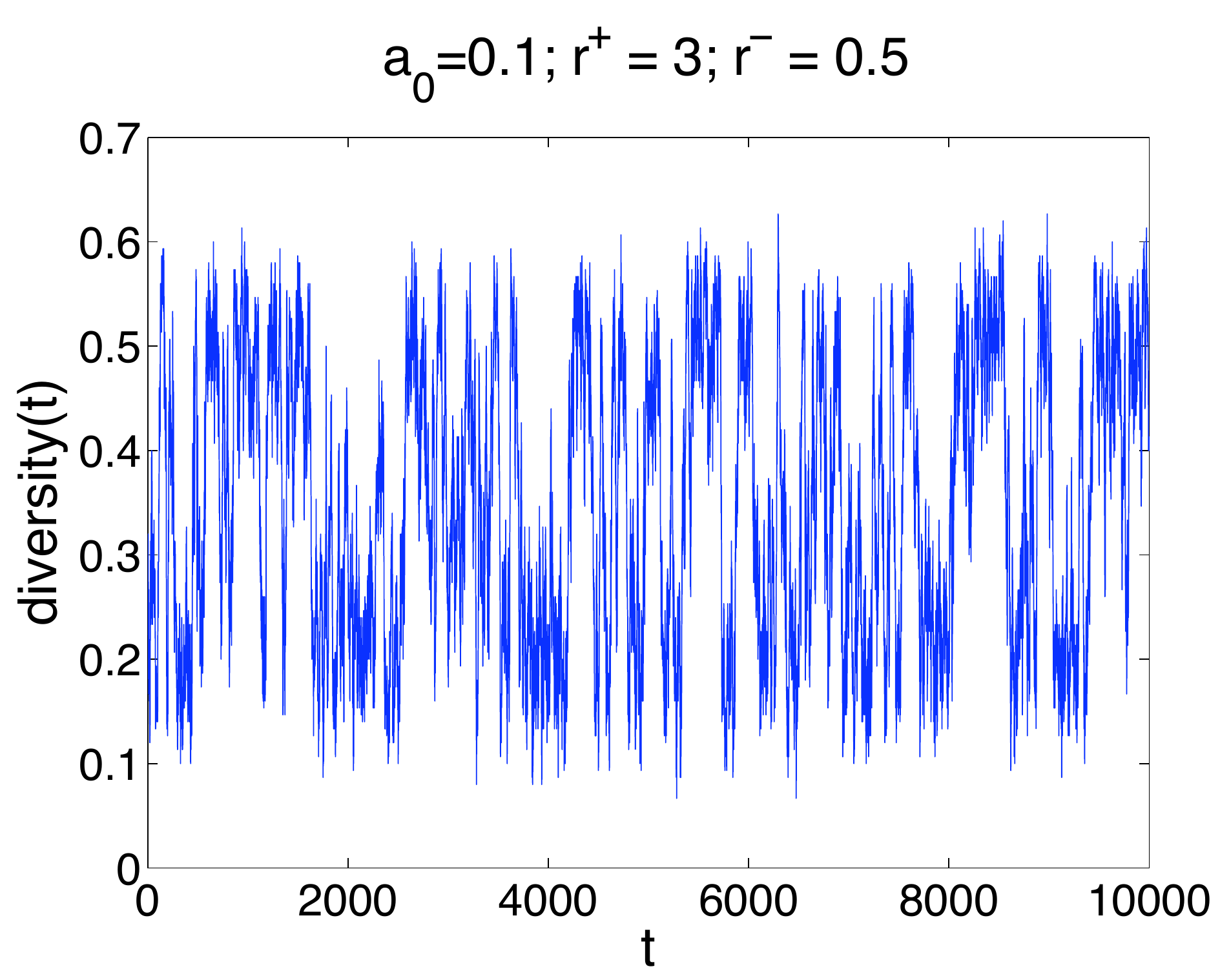}&
\includegraphics[width=5.4cm]{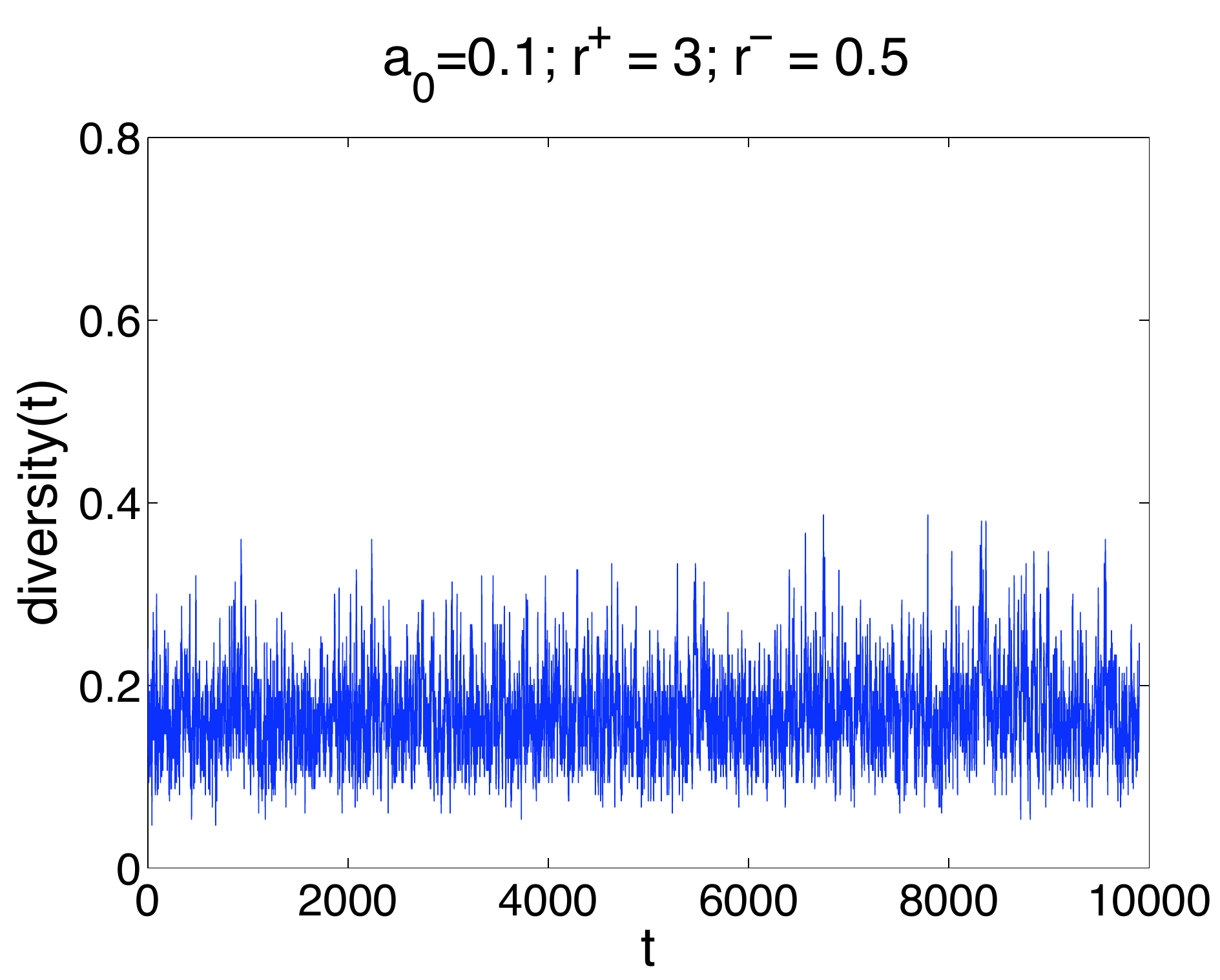}&
\includegraphics[width=5.1cm]{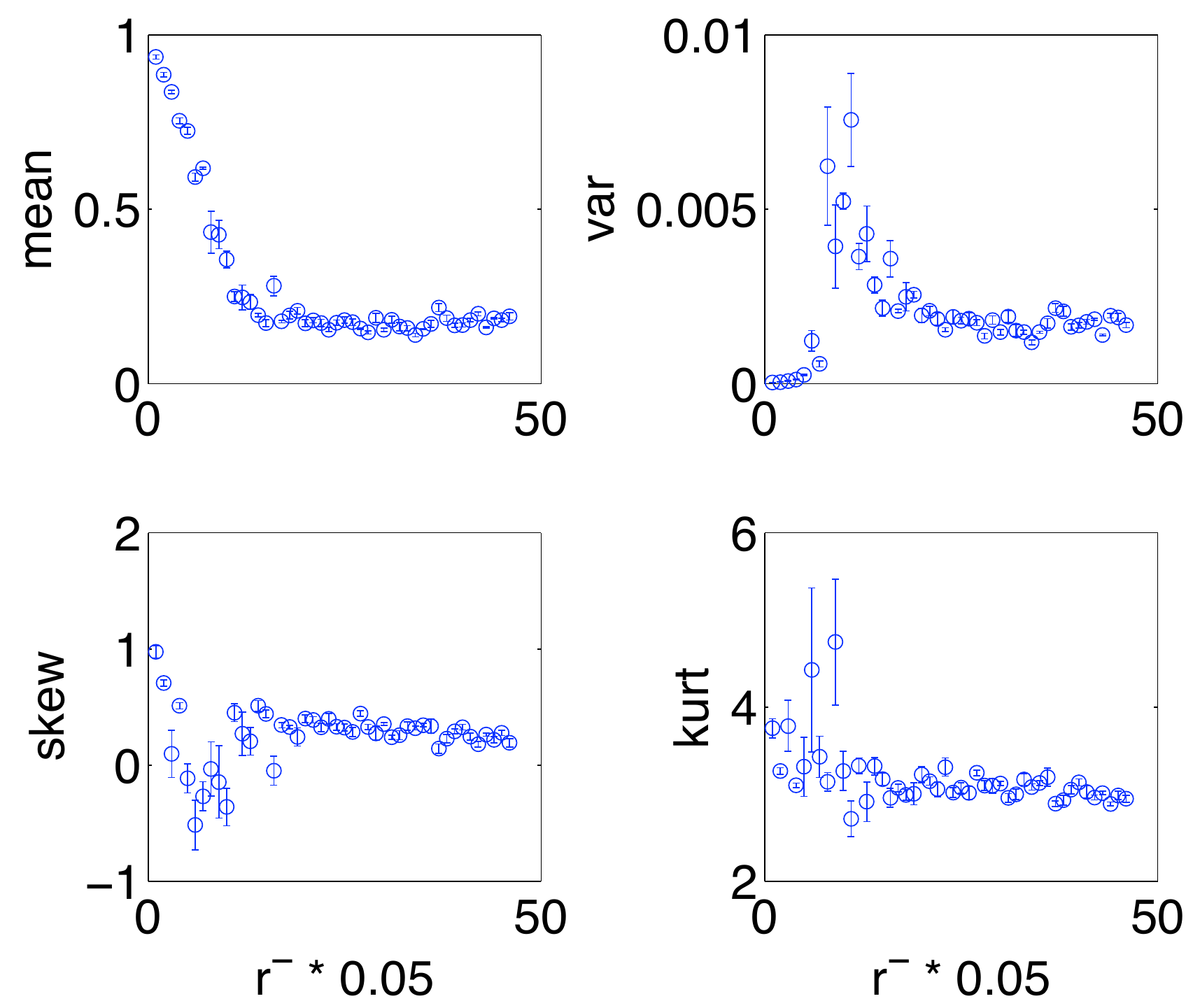}
\end{tabular}
\caption{
Two different runs with the same parameters 
%($t$ is  time and $a(t)$  diversity) 
for two different random seeds of the rule table $\alpha$.
One exhibits meta-bi-stable dynamics (left) the other low-diversity dynamics (center).
In both cases the suppression factor is $\mu=1.5$ and the sustain rate $\lambda=0.08$. The
initial condition $x(0)$ was protected. Statistics for the moments of the diversity dynamics for 
$r^+=3$ and $a_0=0.1$ fixed is shown based 
on 25 runs with different random topologies per $r^-$ (right).
}
\label{fig5}      
\end{figure*}

\subsection{Results}

\begin{figure*}[t]
\includegraphics[width=8cm]{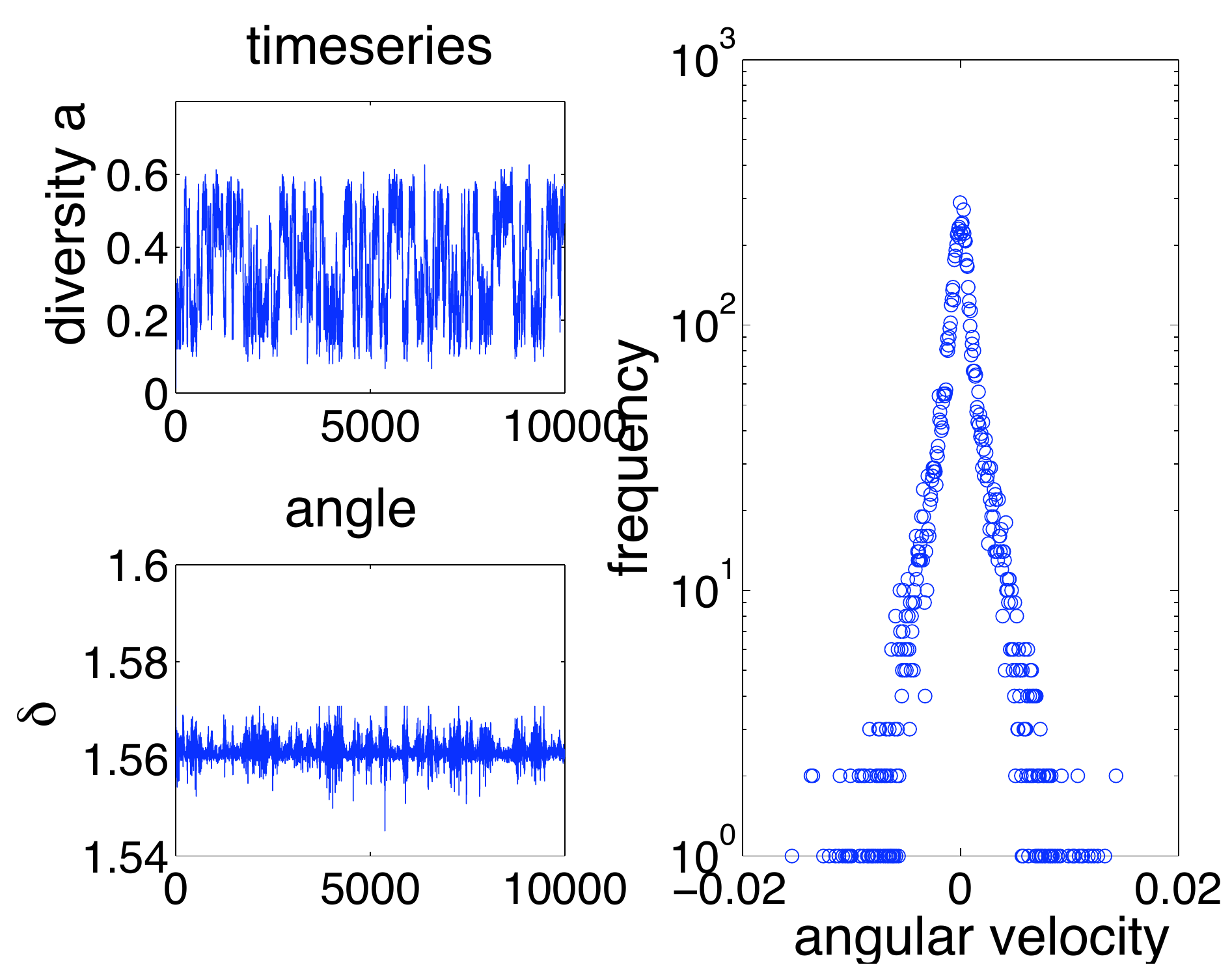}
\caption{ 
Diversity (left upper panel) and angular velocity (left lower panel) 
over time in a particular run displaying meta-bi-stable dynamics. 
The angular velocity distribution is shown in the right plot.
}
\label{fig6}      
\end{figure*}

Most importantly our simple model is capable to produce meta-bi-stable
dynamics, or punctuated equilibrium as called in biology. 
In Fig. \ref{fig5} two runs for identical parameter settings (left  and 
center) are shown. 
The two runs with identical dynamical parameters show meta-bi-stable dynamics in the
one case and sub-critical dynamics in the other. The only difference appears 
in the random realization of the rule table $\alpha$. 
The role of the parameters $\mu$ (relative suppressor strength) and $\lambda$
(sustain rate) are found to have only a minor influence on the qualitative behavior of the dynamics.
In the right image of Fig. \ref{fig5} the first  moments of the diversity timeseries  are
shown as a function of $r^-$, for all other parameters fixed.
We see that the mean diversity decreases with increasing $r^-$ just as we
may expect from Eq. (\ref{suppression-magnitude}),
while the variance is  below a critical $r^-_{crit}$ where the variance
has a peak and then decays for $r^->r^-_{crit}$.
The simulation data we have analyzed indicates that this behavior is widely independent
of the system size $d$, which was typically between 100 and 1000.
We  compared the characteristics of the diversity 
dynamics with direct numerical  solutions of the catalytic-network equation 
Eq. (\ref{networkeq}) and found qualitatively comparable behavior on small systems. 
In case the time-evolution of the system-diversity is meta-bi-stable, the
system seems to be critical. 
In Fig. (\ref{fig6}) the clearly non-Gaussian angular velocity distribution of $\delta$
is shown.

\section{Discussion}

We have proposed a simple model of evolution that is based on a set of
three relevant parameters $r^+$, $r^-$ and $a_0$.
No anthropomorphic {\em a posteriori} concepts like fitness or ecological 
niches are necessary in the model, yet it  is capable of explaining punctuated equilibrium 
in diversity timeseries, which is the most striking feature in  experimental data, 
such as e.g. fossil records. The
creative processes are modeled as a consequence of constructive interactions between pairs of
entities while selection happens implicitly through suppressors, i.e. the existence of certain 
elements prevents (suppresses) the existence of others.
A posteriori it becomes of course possible to interpret fitness and niches in terms of
the interaction topology. However, the model indicates that fitness is maybe  more a
property of the collective state of a system than a property of the individual
entities. 
When the state of the system is such that suppression is minimized and
creativity maximized the whole system thrives while
in case too many suppressors are supported the whole system breaks down in a
'creative gale of destruction' \cite{schumpeter}.  

Most remarkably, already the simple model we have implemented  displays
 meta-bi-stable dynamics -- which is exactly what we would expect to
find -- for a wide range of $r^-$, i.e. for $r^+>r^+_{crit}$,
$a_0 > a_{crit}(r^+)$. One may conclude from this that there
is a non-vanishing probability to sample evolutionary systems with
meta-bi-stable evolution
by pure chance. 

The behavior of mean and variance indicate that when the
system is prepared critically ($a_0 > a_{crit}$) then for small
$r^-$ the system will behave supra-critical and approaches a plateau where
it stays and the variance is small. When $r^-$ increases we find an onset
of meta-bistable evolution. The probability for this
behavior seems to be maximal at the critical value of $r^-$ where the
variance is maximal. For larger $r^-$ the decreasing variance 
indicates both a decreasing probability to find meta-bi-stable dynamics and
a decreasing amplitude of the high-diversity
plateau. The meta-bi-stable scenarios seem to be critical as indicated by the
non-Gaussian distribution of the angular speed $\delta$ of Eq. (\ref{radspeed}).

Finally, since the parameters $a_0$, $r^+$, and $r^-$ are insufficient to fully 
determine whether the dynamics resulting from some random topology $\alpha$
of the catalytic network will be meta-bi-stable, sub-critical, or
supra-critical, this set of parameters is incomplete. 
The occurrence of meta-bi-stability or punctuated equilibrium 
has to be associated with topological properties of the
catalytic network $\alpha$. These properties, such as  auto-catalytic cycles, 
have a non vanishing chance to be randomly sampled.
It will be a fascinating task to relate the   statistics of occurrence 
of such topological structures with the observation of punctuated equilibrium. 
\\
\\
This work was supported by Austrian Science Fund FWF Projects P17621 and P19132.
We thank the organizers of SigmaPhi 2008 for the splendid conference, as well as SFI for its hospitality.

%%%%%%%%%%%%%%%%%%%%%%%%%%%%%%%%%%%%%%%%%%%%%%%%%%%%%%%%%%%%%%%%%%%%%%%%%%
%%%%%%%%%%%%%%%%%%%%%%%%%%%%%%%%%%%%%%%%%%%%%%%%%%%%%%%%%%%%%%%%%%%%%%%%%%
%%%%%%%%%%%%%%%%%%%%%%%%%%%%%%%%%%%%%%%%%%%%%%%%%%%%%%%%%%%%%%%%%%%%%%%%%%
\end{document}